\documentclass{aastex}
\usepackage{emulateapj5}

\shorttitle{Warm Gas Accretion}

\begin{document}
\DeclareGraphicsExtensions{.pdf,.gif,.jpg,.eps}

\title{Gas Accretion is Dominated by Warm Ionized Gas in Milky Way-Mass Galaxies at \lowercase{$z \sim 0$}}

\author{M.~Ryan~Joung, Mary~E.~Putman, Greg~L.~Bryan, Ximena~Fern{\'a}ndez, and J.~E.~G.~Peek\altaffilmark{1}}
\affil{Department of Astronomy, Columbia University, 550 West 120th Street, New York, NY~10027; moo@astro.columbia.edu}
\altaffiltext{1}{Hubble Fellow.}

\begin{abstract}
We perform high-resolution hydrodynamic simulations of a Milky Way-mass galaxy in a fully cosmological setting using the adaptive mesh refinement code, {\it Enzo}, and study the kinematics of gas in the simulated galactic halo.  We find that 
the gas inflow occurs mostly along filamentary structures in the halo.  The warm-hot ($10^5$ K $< T <$ $10^6$ K) and hot ($T > 10^6$ K) ionized gases are found to dominate the overall mass accretion in the system (with $\dot{M} = 3$--5 M$_{\odot}$ yr$^{-1}$) over a large range of distances, extending from the virial radius to the vicinity of the disk.  
Most of the inflowing gas (by mass) does not cool, and the small fraction that manages to cool does so primarily close to the galaxy ($R \lesssim 20$ kpc), perhaps comprising the neutral gas that may be detectable as, e.g., high-velocity clouds.  The neutral clouds 
are embedded within larger, accreting filamentary flows, and represent only a small fraction of the total mass inflow rate.  The inflowing gas has relatively low metallicity ($Z/Z_{\odot} < 0.2$).  The outer layers of the filamentary inflows are heated due to compression as they approach the disk.  
In addition to the inflow, we find high-velocity, metal-enriched outflows of hot gas driven by supernova feedback. 
Our results are consistent with observations of halo gas at low $z$.
\end{abstract}


\section{Introduction}

Chemical evolution models and analysis of the color magnitude diagram of the Hipparcos dataset indicate the Milky Way has been forming stars at a nearly constant, yet slowly declining, rate of 1--3 $M_{\odot}$ yr$^{-1}$ over the past several gigayears \citep{hernandez00, chiappini01, chiappini03, 
fuchs09, chomiuk11}.  
In addition, a continuous supply of low metallicity gas coming in at a slightly lower rate is needed to account for the metallicity of the long-lived Galactic stars \citep[the G-dwarf problem; see][for a review]{tosi96}.  
The source of this fuel has been the subject of decades of research \citep[see][for a review]{putman12}.

High-velocity clouds (HVCs) detected in H{\sc~I} surveys have long been suspected as the source of the star formation fuel \citep{wakker97, putman03}.  Now that distances to most large complexes are known, the mass accretion rate from HVC complexes can be estimated \citep{wakker01, wakker07, wakker08, thom08, putman12}.  The resulting values, however, are too low by about an order of magnitude compared to the recent Galactic star formation rate. 
In addition, observations of nearby, star forming spirals often show a very limited amount of HI gas in their halos \citep{sancisi08, heald11}, again suggesting the HI reservoir in halos may not be the dominant fueling source.

Recently there have been suggestions that extraplanar ionized gas may be responsible for maintaining star formation in the Milky Way \citep{lehner11, putman09}, as well as in other galaxies at intermediate and low redshifts \citep{bauermeister10}.  Hydrodynamic simulations that model the mixing and recooling of cold clouds at the disk-halo interface \citep{heitsch09} and the H$\alpha$ emission along the Magellanic Stream \citep{blandhawthorn07} also pointed out the potential significance of the influx of warm ionized gas.    When combining these results with other findings of substantial amounts of ionized gas in the halos of star forming galaxies at higher redshifts \citep{tumlinson11, tripp11}, it is evident that 
a self-consistent dynamical model is necessary to connect galaxy fueling mechanisms with the various observational constraints.

In this paper, we describe a high-resolution cosmological simulation of a Milky Way-mass disk galaxy using an adaptive mesh refinement (AMR) code and present its key
features in terms of the 
thermal and kinematic distribution of gas in such a halo.  The high mass resolution ($m_{DM}$ and $m_* \approx 10^5 M_{\odot}$) and spatial resolution (136--272 pc comoving or better at all times) employed in the simulation allow us to study and track the spatial and kinematical distribution of the multiphase gas in the halo in great detail.  We describe the simulation in \S~\ref{method}.  The results are presented in \S~\ref{results}, with the emphasis placed on identifying the gas components responsible for inflow onto the galaxy.  Finally, we 
examine the evolution of the gas in filamentary flows in the simulation and present a new scenario for gas accretion onto Milky Way-sized galaxies 
in \S~\ref{discuss}.

\section{Method}
\label{method}

We perform simulations with Enzo, an Eulerian hydrodynamics code with AMR capability (Bryan 1999; Norman \& Bryan 1999; O'Shea et al. 2004).  It solves the Euler equations using the piecewise-parabolic method (PPM; Colella \& Woodward 1984) or the solver used in Zeus (Stone \& Norman 1992) to handle compressible flows with shocks; we used the latter primarily for numerical stability.

First, we ran a low-resolution simulation with a periodic box of $L =$ 25 $h^{-1}$ Mpc comoving on a side with cosmological parameters consistent with WMAP5: ($\Omega_m$, $\Omega_{\Lambda}$, $\Omega_b$, $h$, $\sigma_8$, $n_s$) $=$ (0.279, 0.721, 0.046, 0.70, 0.82, 0.96).  We identified Local Group-like volumes by using criteria based on the halo mass (mass range 1--2 $\times$ 10$^{12}$ M$_{\odot}$), the mean density (0.60--1.0 times the mean density of the universe) and the relatively low velocity dispersion of the halos ($<$ 200 km s$^{-1}$) identified within 5 $h^{-1}$ Mpc of a given galaxy.  We identified four such halos.  Then we performed a resimulation for one of the four halos using the multimass initialization technique with four nested levels (five including the root grid), achieving $m_{DM} = 1.7 \times 10^5$ $M_{\odot}$, within a ($\sim$5 $h^{-1}$ Mpc)$^3$ subvolume.  The selected galaxy has a halo mass of $1.4 \times 10^{12}$ $M_{\odot}$ at $z=0$ and so contains over 8.2 million dark matter particles within the virial radius.  With a maximum of 10 levels of refinement, the maximum spatial resolution stays at 136--272 pc comoving at all times.  Results from the same simulation were discussed in Fern{\'a}ndez et al. (2012), particularly in the context of H{\sc~I} gas.

The simulation includes metallicity-dependent cooling extended down to 10 K (Dalgarno \& McCray 1972), 
metagalactic UV background, shielding of UV radiation by neutral hydrogen, and a diffuse form of photoelectric heating (Abbott 1982; Joung et al. 2009).   The code simultaneously solves a complex chemical network involving multiple species (e.g., H~{\sc I}, H~{\sc II}, H$_2$, He~{\sc I}, He~{\sc II}, He~{\sc III}, e$^-$) and metal densities explicitly.

Star formation and stellar feedback, with a minimum initial star particle mass of $m_* = 1.0 \times 10^5$ $M_{\odot}$, are also included.  Star particles are created in cells that satisfy the following two criteria:  $\rho > \rho_{SF}$ and a violation of the Truelove criterion (Truelove et al. 1997).  The star formation efficiency (i.e., the fraction of gaseous mass converted to stars per dynamical time) is 0.03 (e.g., Krumholz \& Tan 2007).  
Supernovae feedback is modeled following Cen et al. (2005), with the fraction of the stellar rest-mass energy returned to the gas as thermal energy, $e_{SN} = 10^{-5}$.  Feedback energy and ejected metals are distributed into 27 local cells centered at the star particle in question, weighted by the specific volume of the cell.  The temporal release of metal-enriched gas and thermal energy at time $t$ has the following form: $f(t, t_i, t_*) = (1/t_*)[(t-t_i)/t_*] \exp[-(t-t_i)/t_*]$, where $t_i$ is the formation time of a given star particle, and $t_* =$ max($t_{\rm dyn}$, $3 \times 10^6$ yr) where $t_{\rm dyn} = \sqrt{3 \pi / (32 G \rho_{\rm tot})}$ is the dynamical time of the gas from which the star particle formed.  The metal enrichment inside galaxies and in the intergalactic medium (IGM) is followed self-consistently in a spatially resolved fashion.

\section{Results}
\label{results}

We extracted a spherical volume from the simulation output that extends to the galaxy's virial radius (250 kpc) at a uniform spatial resolution of 1.09 kpc/cell.  
In order to examine finer structures, the volume inside 20 kpc in radius was extracted at a higher resolution of 0.272 kpc/cell, the maximum spatial resolution of the simulation, and this replaced the inner volume of the larger sphere.  
In order to focus on gas accretion in the halo, the cylindrical region defined by $|R| \le 18$ kpc and $|z| \le 2$ kpc whose symmetry axis coincides with the rotation axis of the simulated disk was removed from this analysis.  Hence, the resulting quantities reflect the properties of the halo region only.
We report on our analysis of the simulation result at $z=0$, unless otherwise specified.  The evolution of H{\sc~I} gas in the halo at low redshifts ($z \le 0.5$) was studied in detail by \citet{fernandez12}.

We find that 70\% of the mass influx is concentrated in $\sim$17\% of the surface area over a large range of radii.  This implies that the gas inflow occurs along continuous, filamentary structures.  We find three main filaments of warm gas that feed the galaxy.  Further details on the spatial and kinematic properties of these warm filamentary flows will be reported in a forthcoming paper.  

\subsection{Gas Inflow Velocities}
\label{vinfl}

We examine the distribution of radial velocities of the halo gas.  
The systemic velocity of the galaxy, i.e,, the center-of-mass velocity of the dense ($n \ge 0.1$ cm$^{-3}$) cells in the disk, was subtracted from all cells, to focus on the {\it relative} motion of the halo gas with respect to the galaxy itself.

\begin{figure}
\vspace{-5mm} \hspace{30mm} \includegraphics[scale=1.2]{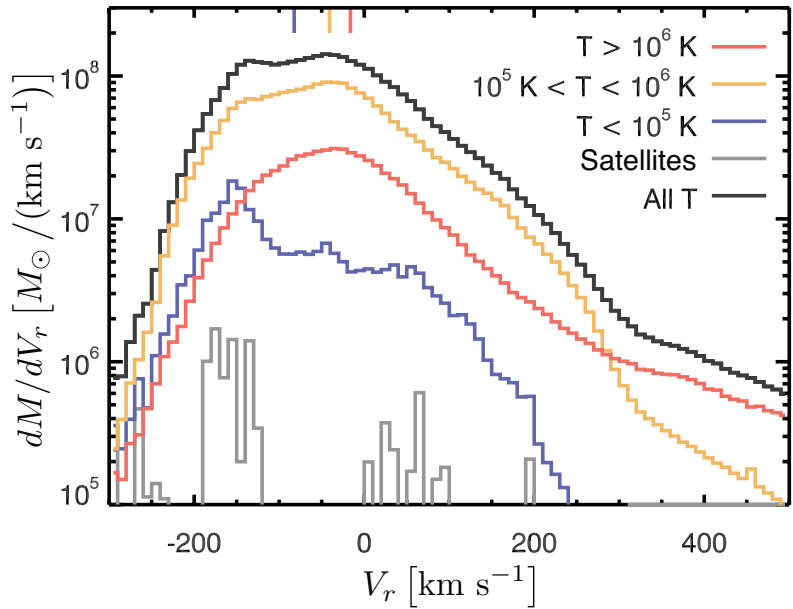}
\caption{Mass contained in unit velocity interval as a function of the radial velocity of gas in three different temperature ranges, as described in the text, at $z=0$: cold ($T < 10^5$ K; blue), warm-hot (10$^5$ K $< T <$ 10$^6$ K; yellow), and hot ($T > 10^6$ K; red).  Positive (negative) velocities correspond to outflows (inflows).  The black solid line is the sum of the three solid curves in color, while the grey dotted line indicates the amount of cold gas contained within 10 kpc of the four gas-rich satellites identified within the virial radius of the simulated host halo.  Three of the four satellites have $|v_r| > 100$ km s$^{-1}$.
The warm-hot gas dominates the mass over almost the entire range of radial velocities, except at the highest values (i.e., $v_r \gtrsim 300$ km s$^{-1}$), where the hot outflowing gas contributes most significantly with $\sim$10$^8$ M$_{\odot}$ of mass.  The mass-weighted mean radial velocities are $-82$, $-41$, and $-16$ km s$^{-1}$ for the cold, warm-hot, and hot components, respectively, and are shown by vertical lines at the top of the plot.
\label{vr}}
\end{figure}

Figure \ref{vr} displays the radial velocity distribution of gas in various temperature ranges.  It shows the curves representing the amount of mass per unit velocity interval in three different temperature ranges, plotted against the radial velocity. 
The three temperature ranges were selected to be cold ($T < 10^5$ K, blue), representative of H{\sc~I} \& H$\alpha$ emission and Ly$\alpha$ \& Mg{\sc~II} absorbers;  warm-hot (10$^5$ K $< T <$ 10$^6$ K, yellow), representative of C{\sc~IV} and O{\sc~VI} absorbers;  and hot ($T > 10^6$ K, red), representative of higher level ions such as O{\sc~VII} and O{\sc~VIII} as well as X-ray emission.  These definitions are used in Figures \ref{maccr} and \ref{maccrZ} as well.  The gas associated with the last component has densities that are usually too low to be detected in current observations, except for the region close to the disk.  The black curve is the sum of the three solid curves mentioned above. 

\placefigure{vr}

The warm-hot gas dominates the mass over almost the entire range of radial velocities.  
Although the hot gas occupies a significant volume fraction, it does not dominate the mass because of the low densities.  
The only exception is at the highest radial velocities ($v_r \gtrsim 300$ km s$^{-1}$), where the hot outflowing gas contributes 
$\sim$10$^8$ M$_{\odot}$.  

The grey histogram shows the amount of cold gas contained within 10 kpc of the four gas-rich satellites identified within the virial radius of the simulated host halo (see below for more details).  We picked the radius of 10 kpc because it is at least 40\% (and up to 100\%) of the virial radii of the satellite subhalos, and so the bulk of the cold gas should reside within this volume, unless it was previously ejected or stripped away \citep[see][]{fernandez12}.  Three of the four satellites have $|v_r| > 100$ km s$^{-1}$, suggesting that at least part of the cold gas with extreme velocities must be associated with gas contained within or stripped recently from the satellite galaxies.

The mean radial velocity increases with gas temperature from more negative to less negative velocities.  We find that the cold gas has more negative inflow velocities (the mass-weighted mean radial velocity $\langle v_r \rangle = -82$ km s$^{-1}$) than the warm-hot and hot gases ($\langle v_r \rangle = -41$ and $-16$ km s$^{-1}$, respectively).  These values are marked by vertical lines at the top of the figure.  The inflowing velocities of cold gas are consistent with observations of HVCs, although we leave the details of neutral gas structure, projection effects from the position and velocity of the Sun, and obscuration by Galactic disk gas to future work.  Note that the radial velocities alone do not tell us which phase is primarily responsible for the gas inflow;  we must examine the mass flux in the radial direction to answer that.

\subsection{Mass Accretion Rate}

\begin{figure}
\vspace{-5mm} \hspace{35mm} \includegraphics[scale=1.0,angle=0]{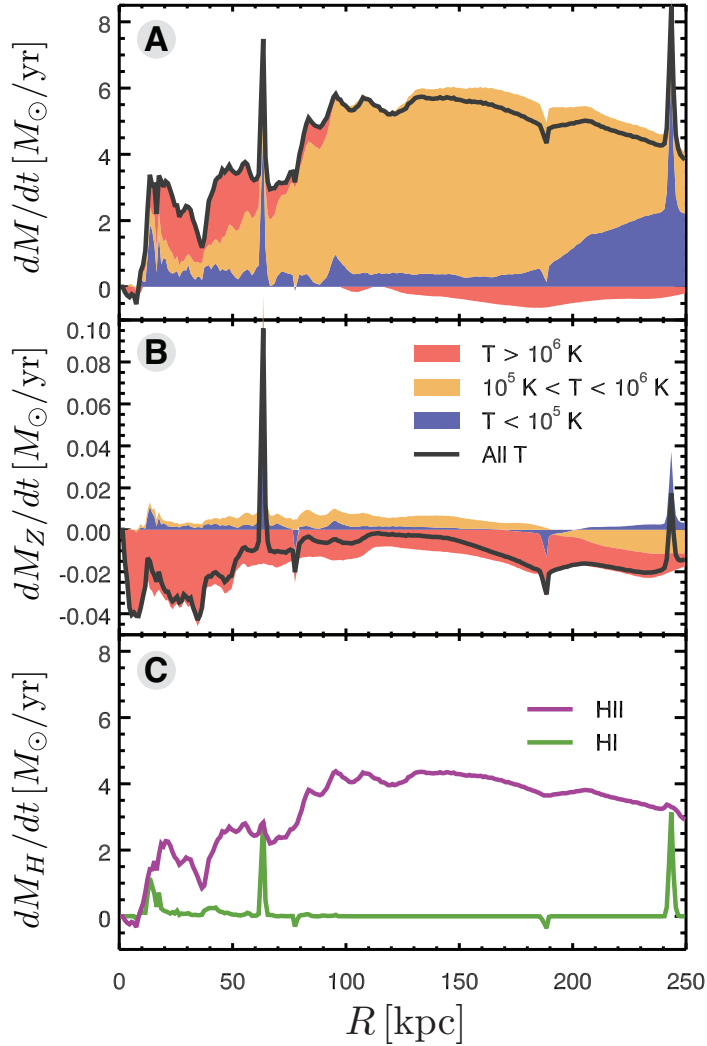}
\caption{Net mass accretion rate of ($a$) all gas, ($b$) metals, and ($c$) hydrogen gas as a function of the galactocentric distance.  The cylindrical region containing the disk has been removed from the analysis, as described in the text.  In the top and middle panels ($a$ and $b$), the blue, yellow, and red areas indicate the appropriate quantity for the cold, warm-hot, and hot components, respectively.  In this figure and Figure \ref{maccrZ}, the sign of the mass flow rate is defined such that inflows will correspond to positive values.  In the top panel ($a$), the dominant component for the gas accretion switches from cold to warm-hot gas at $R \approx 240$ kpc.  Inside of $R \approx 50$ kpc, the hot gas dominates the influx, although in the region closest to the disk ($R \lesssim 20$ kpc), all three components contribute about equally.  In contrast to the gas, metals are outflowing (i.e., $dM_Z/dt < 0$ in our convention) from the system at almost all radii ($b$).  Despite the low overall mass fraction, the hot component (red), which is highly metal-enriched, accounts for the bulk of the metal outflow.  The bottom panel ($c$) shows that the accretion of the ionized hydrogen gas (H{\sc~II}, purple) dominates over that of the neutral hydrogen gas (H{\sc~I}, green).  The sharp features at 13.5, 63, 78, 188 and 243 kpc that can be identified in all the panels -- but mostly distinctively in the H{\sc~I} accretion rate -- are from gas associated with the four gas-rich satellite galaxies present within the virial radius at $z=0$.
\label{maccr}}
\end{figure}

Figure \ref{maccr} shows the mass accretion rate of gas as a function of the galacto-centric distance.  To calculate the mass accretion rate in thin spherical shells centered on the galaxy, we used a formula from \citet{peek08}:
\begin{equation}
\dot{M}(R) = \sum_{i=1}^{n(R)} \, \frac{M_i \, \mbox{\boldmath$V_i$} \cdot (-\hat{r}_i)}{dR} \; ,
\end{equation}
where $M_i$ is the gas mass in the $i^{th}$ cell in a given spherical shell, $\mbox{\boldmath$V_i$}$ is the velocity vector of that cell, $\hat{r}_i$ is the radial unit vector, and $dR$ is the thickness of the spherical shell.  Note that this formula gives the mass accretion rate for gas contained in each spherical shell in units of $M_{\odot}$ yr$^{-1}$.  

\placefigure{maccr}

Plotted in Figure \ref{maccr} are the net (i.e., inflow minus outflow) mass accretion rates of all gas ($a$), of the metals ($b$), and of the neutral and ionized hydrogen ($c$).  In each panel, the mass accretion rates were divided into the three temperature ranges defined in \S\ref{vinfl}: cold (blue), warm-hot (yellow), and hot (red).  

The net mass accretion rate, 3--5 $M_{\odot}$ yr$^{-1}$, at all radii is comparable to the star formation rate of the simulated galaxy at $z \approx 0$ ($\sim$5 $M_{\odot}$ yr$^{-1}$).\footnote{As in most other cosmological simulations run to date, the stellar mass ($\sim$$1.9 \times 10^{11}$ $M_{\odot}$) of our simulated galaxy is too large and too centrally concentrated;  the associated star formation rate at low redshifts is also too high by factors of 2--3.}  The fluctuation in the mass accretion rate is expected from the clumpy and stochastic nature of the accreting mechanisms.  
The amount of neutral gas mass increases at small distances from the galaxy \citep[Figure \ref{maccr}$c$;  see also Figure 1 in][]{fernandez12}, implying some cooling and condensation of gas close to the disk due to increased background pressure.  

The primary result of this paper is shown in Figure \ref{maccr}$a$ and \ref{maccr}$c$ displaying the mass accretion rates;  the overall gas accretion is dominated by warm-hot ionized gas, rather than cold neutral gas, at almost all radii.  The bottom panel (Fig. \ref{maccr}$c$), which shows the accretion rate of hydrogen gas, demonstrates that the ionized gas is responsible for most of the mass influx with $dM_{HII}/dt \approx$ 2--4 $M_{\odot}$ yr$^{-1}$, while the neutral gas accounts for only 0.1--0.3 $M_{\odot}$ yr$^{-1}$ (excluding satellites).  This is due, in part, to the fact that the filamentary flows responsible for roughly half of the neutral gas in the halo \citep{fernandez12} are associated with temperatures between 10$^4$ and 10$^{5.5}$ K and are mostly ionized.  Closer to the disk ($R \lesssim 100$ kpc), warm-hot gas is gradually heated, and the accretion of hot gas becomes increasingly important (but see the caveat in \S\ref{discuss}). 

The four sharp features in the H{\sc~I} accretion rate at $R \approx$ 63, 78, 188, and 243 kpc correspond to the four gas-rich satellites (with $N_{HI} \ge 10^{16}$ cm$^{-2}$) found within the virial radius at $z=0$.  The negative values correspond to those satellites moving away from the galaxy.  
The feature peaked at $R \approx$ 13.5 kpc is also associated with one of the four satellite galaxies, although in this case the bulk of its mass may come from condensation of gas stripped from the satellite galaxy, which is at $R = 78$ kpc and moving away from the host galaxy at $z=0$ \citep[``S19" in][]{fernandez12}.  This and other H{\sc~I} features close to the disk may indicate cooling of the inflowing halo gas at the disk-halo interface.  We will investigate this issue in more detail in the future.  
Note that the UV radiation from young stars in the disk, which is not included in the simulation, may photoionize some of the neutral gas in the halo.  Hence, our calculated H{\sc~I} mass is an upper limit.

\begin{figure}
\vspace{-10mm} \hspace{39mm} \includegraphics[scale=1.0,angle=0]{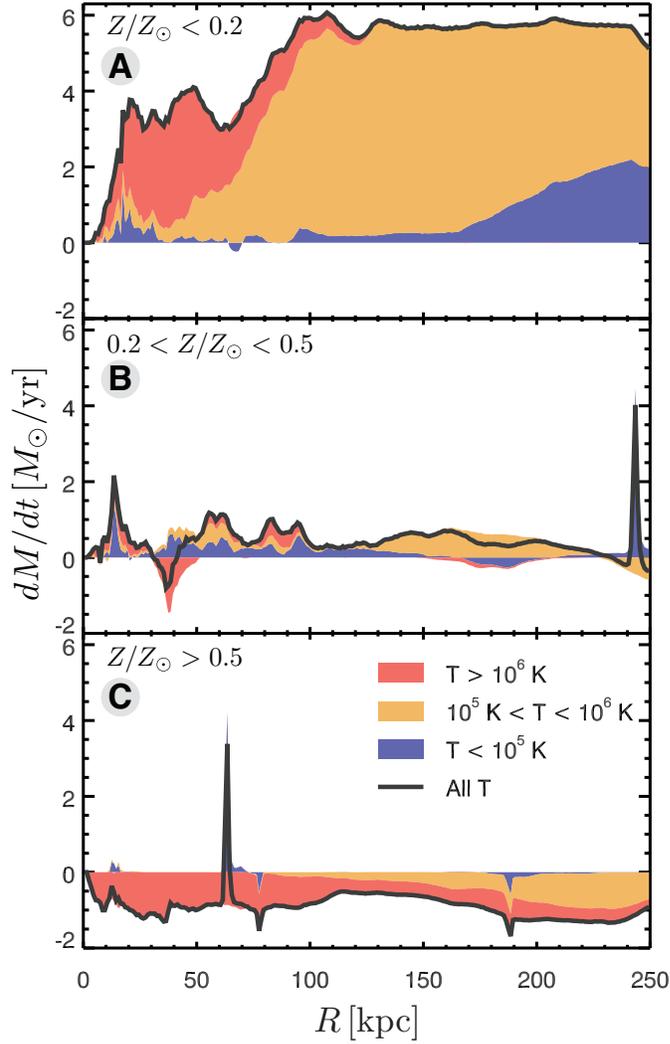}
\caption{Similar to Figure \ref{maccr} but each panel corresponds to one of the three metallicity ranges:  ($a$) low ($Z/Z_{\odot} \le 0.2$), ($b$) intermediate ($0.2 < Z/Z_{\odot} \le 0.5$), and ($c$) high ($Z/Z_{\odot} > 0.5$).  The black solid curves in the three panels represent the accretion rate of gas in each of these metallicity bins.  To display the temperature distribution of the inflowing/outflowing gas, the curves are further divided into three temperature bins using the same colors as in, e.g., Figure \ref{maccr}$a$.  The low-metallicity gas dominates the inflow in all the phases.  The bottom panel ($c$) shows a clear gas outflow of high-metallicity gas at the rate of $\sim$1 $M_{\odot}$ yr$^{-1}$.  The temperature of the outflowing gas decreases gradually from hot at small radii ($R \lesssim 120$ kpc) to warm-hot at larger radii ($R \gtrsim 120$ kpc), presumably due to adiabatic expansion.
\label{maccrZ}}
\end{figure}

\placefigure{maccrZ}

In Figure \ref{maccrZ}, each panel corresponds to one of the three metallicity ranges:  ($a$) low ($Z/Z_{\odot} \le 0.2$), ($b$) intermediate ($0.2 < Z/Z_{\odot} \le 0.5$), and ($c$) high ($Z/Z_{\odot} > 0.5$).  The black solid curves in the three panels represent the accretion rate of gas in each of these metallicity bins.  To display the temperature distribution of the inflowing/outflowing gas, the curves are further divided into three temperature bins using the same colors as in, e.g., Figure \ref{maccr}$a$.  The figure demonstrates that it is the low-metallicity gas that dominates the inflow in all the phases.  As the low-metallicity gas flows in, it makes a smooth transition from cold to warm-hot and then from warm-hot to hot temperatures.

The bottom panel shows a clear gas outflow of high-metallicity gas at the rate of $\sim$1 $M_{\odot}$ yr$^{-1}$.  The temperature of the outflowing gas decreases gradually from hot at small radii ($R \lesssim 100$ kpc; bottom panel) to warm-hot at larger radii (top right panel), presumably due to adiabatic expansion.  The metals are carried in hot outflowing gas, as predicted by previous theoretical work \citep[e.g.,][]{maclow89, strickland00, marcolini05}.  This result is consistent with observations finding highly metal-enriched hot gas in the X-ray \citep[e.g.,][]{strickland07} and in the ultraviolet \citep[e.g.,][]{tripp11}.  Note that the metals have a net outflow rate at almost all radii, although the total gas accretion rate always indicates a net inflow.

\section{Discussion}
\label{discuss}

We showed that the overall gas accretion is dominated by warm-hot ionized gas rather than cold neutral gas in Milky Way-sized galaxies at low redshifts.  
According to Figure \ref{maccr}$a$, the component that dominates the gas accretion changes gradually from cold to warm-hot (at $R \approx 240$ kpc) and then from warm-hot to hot (at $R \approx 50$ kpc), as the distance decreases.  What is responsible for the gradual heating of the inflowing gas?

\begin{figure}
\vspace{-0mm} \hspace{22mm} \includegraphics[scale=1.2,angle=0]{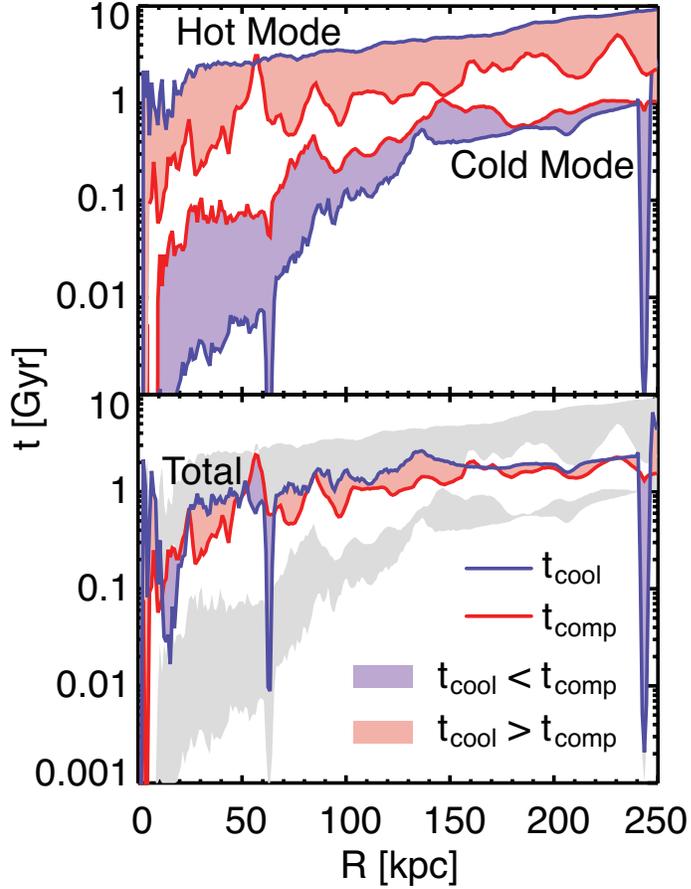}
\caption{{\it (Top panel)} Mean cooling time (blue) and compression time (red) as a function of the galactocentric radius, for the inflowing hot mode gas (upper curves) and cold mode gas only (lower curves).  To compute the mean timescales, the inverse of the appropriate time was weighted by the thermal energy density ($\frac{3}{2} n k T$) in each cell and summed over all inflowing ($v_r < 0$) cells in a given spherical shell.  For the hot mode gas, the mean cooling time is longer than the compression time ($t_{cool} > t_{comp}$, represented by the pink area), leading to net heating.  For the cold mode gas, on the other hand, the mean cooling time is shorter than the compression time ($t_{cool} < t_{comp}$, denoted by the purple area), resulting in net cooling;  note that the difference between the two times increases at small radii.  If the same plot is made for all inflowing gas {\it (bottom panel)}, the two timescales are nearly equal (within a factor of $\sim$2) over the entire range in radius.  Overall, excluding the sites of gas associated with the satellites, the heating should slightly dominate over cooling.  The timescales in the top panel are reproduced in light grey in the bottom panel.
\label{coolcomp}}
\end{figure}

To address this question, we plot in Figure \ref{coolcomp} the cooling time and compression time vs. radius for the ``hot mode" gas (Fig. \ref{coolcomp}, upper curves) and ``cold mode" gas (lower curves), which are hereafter defined as gas with $T > 10^{5.5}$ K and with $T \le 10^{5.5}$ K, respectively.  This is similar\footnote{In \citet{keres05}, the hot mode accretion and cold mode accretion were defined based on the maximum temperature attained by a given gas particle.  As we cannot follow the history of gas particles in a grid-based code, in order to define the two modes, we use the temperature at a given time slice instead of the maximum temperature.  For this reason, the mass of the hot mode gas that we compute is a lower limit.  In particular, some fraction of the gas with $T \le 10^{5.5}$ K at large radii will likely be heated further and so should really count as hot mode gas.  On the other hand, we believe that the distinction between the two modes at small radii ($R \lesssim 100$ kpc) is robust.  Also, note that the definitions for the cold mode and hot mode gas are to be distinguished from those for the cold, warm-hot, and hot gas in \S\ref{results}.} to the definition in \citet{keres05}.  To compute the mean timescales, the inverse of the appropriate time was weighted by the thermal energy density ($\frac{3}{2} n k T$) in each cell and summed over all inflowing ($v_r < 0$) cells in a given spherical shell.  In computing the cooling time, the diffuse photoelectric heating rate was also accounted for.

For the hot mode inflowing gas, the cooling time (blue) is longer than the compression time (red) at all radii, which suggests that heating dominates over cooling for this component.  On the other hand, if we repeat the same calculation for the cold mode gas, the cooling time is shorter than the compression time at all radii, leading to net cooling and condensation of the densest parts of the inflowing streams, especially at small radii ($R \lesssim 100$ kpc).  If we make the same plot for all inflowing gas (bottom panel), the two timescales are nearly equal (within a factor of $\sim$2) over the entire range in radius, excluding the sites of gas associated with the satellites.  Overall, the compression time is shorter than the cooling time, so heating should slightly dominate over cooling.

\placefigure{coolcomp}

As the gas flows in, its kinetic energy gets slowly converted to thermal energy due to many weak compressions.  The energetics work out since $v_r \approx 100$ km s$^{-1}$ would correspond to $\sim$10$^6$ K in gas temperature.  The result is consistent with the trend in Figure \ref{vr}, i.e.,  the mean inflow velocity decreases as the gas temperature increases.
 
\begin{figure}
\hspace{-10mm} \includegraphics[scale=0.40,angle=90]{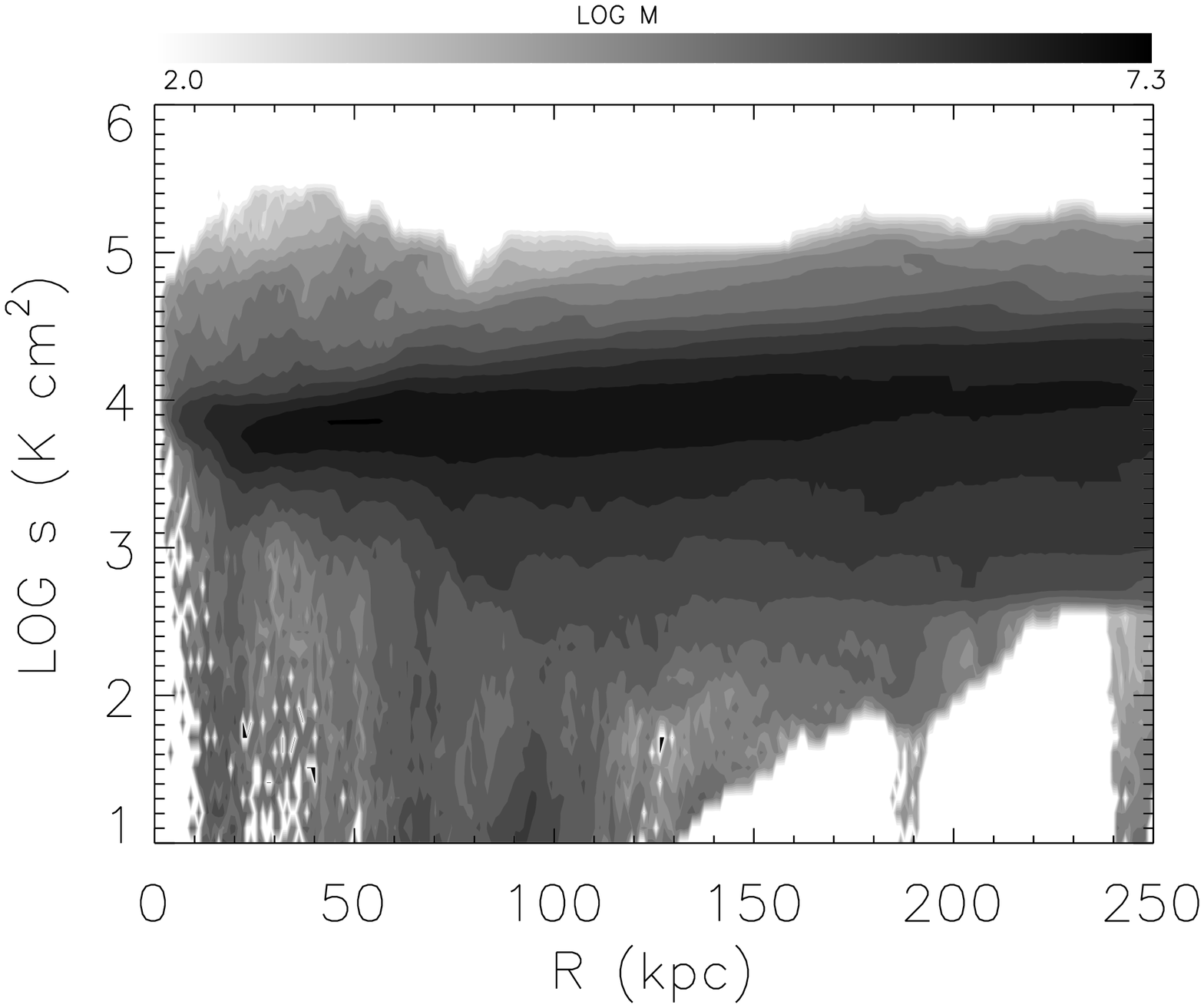}
\hspace{-12mm} \includegraphics[scale=0.40,angle=90]{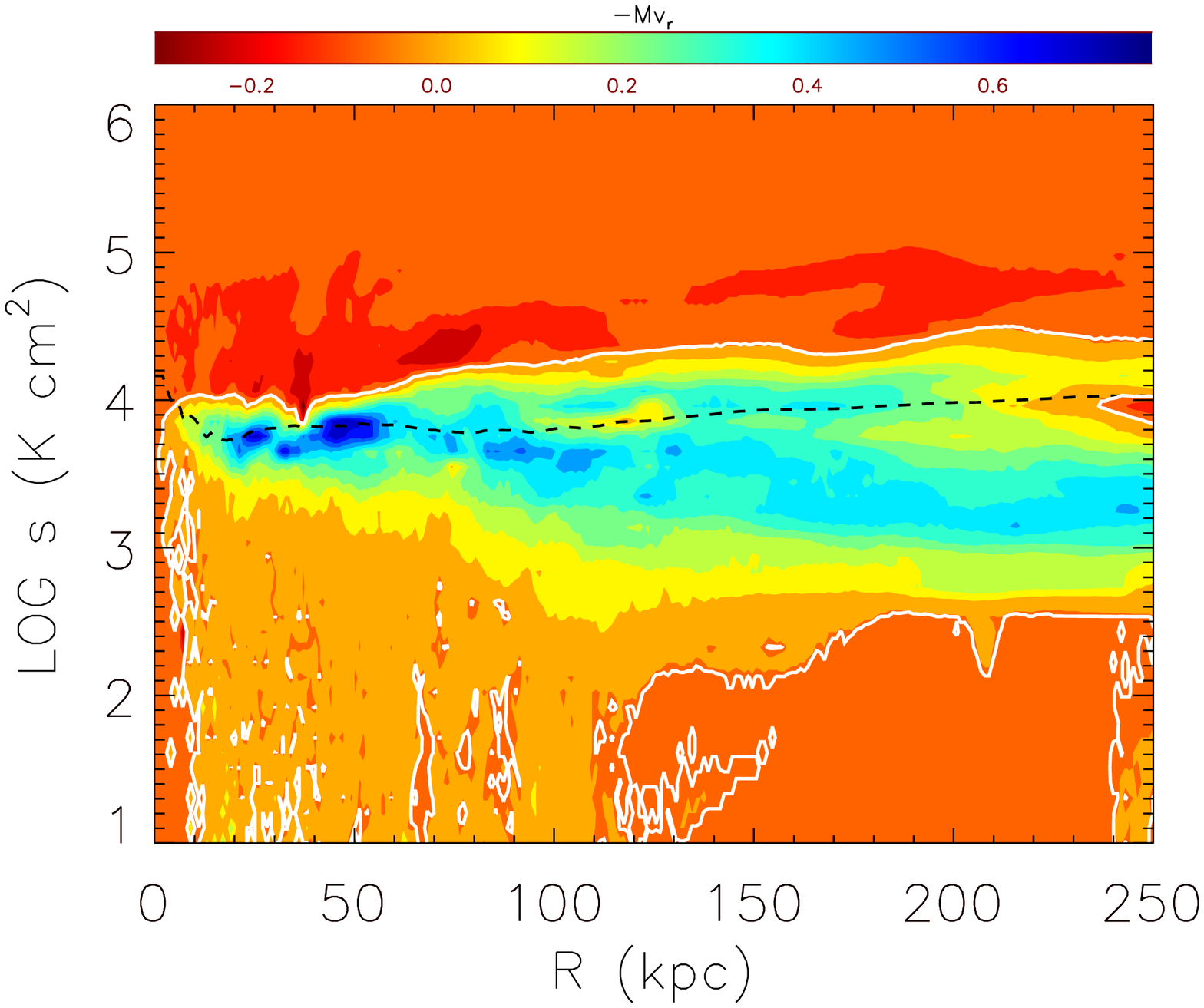}
\caption{Contours showing the distribution of gas mass {\it (left)} and the mass flux {\it (right)} in arbitrary units in the specific entropy vs. distance plane.  
The mean entropy is fairly flat with radius, increasing by only a factor of two from 20 to 250 kpc.  In the right panel, blue and red represent inflowing and outflowing fluxes, respectively.  The white contour represents zero net mass flux, while the dashed curve shows the mass-weighted mean entropy computed from the left panel.  The inflowing gas (blue region in the right panel) has systematically lower entropy than typical gas at that same radius, by a factor of 2--5 at $R \gtrsim 100$ kpc, which is less than the critical cloud overdensity required for cooling (Joung et al. 2012).  Note the ``beard" in the lower left corners of both panels;  they represent the cooling gas in the innermost regions inside the filamentary flows at $R \lesssim 100$ kpc.
\label{contours_entr}}
\end{figure}

Figure \ref{contours_entr} shows the distribution of specific entropy, $s \equiv T \rho^{1-\gamma}$ where $\gamma$ is the adiabatic index, as a function of radius, weighted by gas mass (left panel) and mass flux (right panel).  (Although the correct term for $s$ is `adiabat,' we refer to it as entropy following common convention.)  This shows that the mean entropy is fairly flat with radius, increasing by only a factor of two from 20 to 250 kpc.  In the right panel, blue and red represent inflowing and outflowing fluxes, respectively.  The white contour represents zero net mass flux, while the dashed curve shows the mass-weighted mean entropy computed from the left panel.  Comparing the two panels, we see that the inflowing gas (blue in the right panel) has systematically lower entropy than typical gas at that same radius, by a factor of 2--5 at $R \gtrsim 100$ kpc.  This corresponds to a density enhancement of less than a factor of 10 for the inflowing gas, assuming pressure equilibrium.  This is less than the critical cloud overdensity required for cooling found by \citet{joung12} in idealized simulations, and therefore consistent with the fact that we do not see cooling instabilities at large radii ($R \gtrsim 100$ kpc).

At smaller radii, we do see cooling gas -- in fact, the ``beard" in the lower left region of the two plots represent the gas cooling at $R \lesssim 100$ kpc.  It is not immediately clear how much gas is cooling, but when we rank the individual cells by the mass influx at a given distance, we find that this cooling gas accounts for only the bottom $\sim$10\% of the mass influx.  The cooling gas corresponds to the innermost regions inside the filamentary flows, which have the highest densities and lowest temperatures -- hence the lowest entropies at a given radius.  This gas is responsible for the small H{\sc~I} clouds at $R < 100$ kpc in Figure 2$c$.

The rest of the inflowing gas, with higher specific entropy, has nearly constant entropy as it flows in;  in fact, there appears to be a gradual increase in the entropy due to mixing or weak shocks.  The temperature of this gas increases as it approaches the disk and is compressed.  It might seem surprising that the gas temperature increases as the filaments flow in;  however, it is consistent with Figure \ref{coolcomp}, which shows that the cooling time of this inflowing warm-hot gas is longer than the compression time for the hot mode gas.

Our result is consistent with previous work \citep[e.g.,][]{keres05, keres09, dekel06} that found the broad idea that hot mode accretion dominates the overall gas accretion in Milky Way-mass halos at low $z$.  However, the inflowing filamentary gas is not strongly shocked and has lower temperatures than the rest of the halo gas.
 
\placefigure{contours_entr}

The observed neutral gas clouds are only the tip of a much larger ``iceberg";  H{\sc~I} clouds represent the small densest parts of filamentary flows that are made up of mostly ionized gas, in which they are embedded.  Neutral gas accounts for a significant but not a dominant fraction ($\sim$1/10) of the accretion rate required to explain the current Galactic star formation rate.  This is comparable to the recent estimate of $\sim$0.1 $M_{\odot}$ yr$^{-1}$ from all the HVC complexes \citep{putman12}.  Note that the outer envelope with intermediate velocities will dynamically ``shield" the H{\sc~I} clouds from the (nearly static) ambient medium and decrease the relative velocity, increasing the Kelvin-Helmholtz growth time and hence the cloud lifetimes.  Figure \ref{maccrZ}, which shows the mass accretion rate broken down by temperature for each metallicity range, demonstrates that the low-$Z$ gas dominates the accretion at all radii for all the phases, which supports the filamentary flow origin of the inflowing gas.

The simulation results are consistent with observations of halo gas.  The H{\sc~I} clouds found in galaxy halos are largely within 20 kpc of galactic disks \citep{thom08, wakker08}, while the (largely) ionized component extends throughout the halo \citep{prochaska11, bowen02}.  In addition, the H{\sc~I} component is surrounded by warm and warm-hot gas indicating multiphase flows are present \citep{putman12, sembach03}.  Finally, substantial quantities of inflowing warm gas are consistent with the results of \citet{shull09} and \citet{lehner11}.

We must point out one caveat in our analysis.  The stellar mass of the simulated galaxy is too concentrated in the bulge, and so the gravitational potential well has a slope that is too steep compared to the Milky Way.  For this reason, the heating of the incoming gas was likely overestimated.  However, it will probably introduce only a factor of a few error in gas temperature, and we expect the qualitative results reported in this paper to remain unchanged.

\section{Conclusions}

We analyzed a high-resolution AMR cosmological simulation of a Milky Way-mass galaxy including star formation and supernova feedback, in a fully cosmological setting.  In summary, our key results are:
\begin{itemize}
\item The inflowing gas is filamentary, and the bulk of the inflow is warm-hot (10$^5$ K $< T <$ 10$^6$ K) and ionized.

\item Most of the inflowing gas (by mass) does not cool;  it has nearly constant entropy and so the temperature increases as the gas approaches the center.

\item Some of the inflowing gas does manage to cool (in the innermost regions of the filaments associated with the lowest entropies), but only inside $R \lesssim 100$ kpc, and mostly within $R \lesssim 20$ kpc.

\item The inflowing gas has low metallicity ($Z/Z_{\odot} < 0.2$).

\item The typical inflow velocities are 50--150 km s$^{-1}$ and generally decrease with increasing gas temperature.
\end{itemize} 

These results point to a picture in which filamentary gas flows, driven by the cosmic web, continue to be important in Milky Way-mass galaxies at low redshifts.  This inflow is not ``cold mode accretion" in the sense of \citet{keres05}, since the temperatures typically exceed $10^{5.5}$ K during the passage through the halo, and radiative cooling does not dominate heating.  However, it also does not correspond to classic smooth, hot-mode accretion, and the gas in these filaments do not experience a large entropy jump at the accretion shock.  Instead, this warm-hot filamentary flow may represent a third mode of accretion -- important for galaxies like the Milky Way that are not far beyond the mass and redshift thresholds below which cold-mode accretion dominates.  We suggest two areas for future work:  the fate of the flows as they reach and enter the galactic disk at the disk-halo interface, and an exploration of how mergers, AGN, and feedback may affect the gas while it is still in the IGM.



\end{document}